\documentclass[a4paper]{jpconf}

\usepackage[dvips]{graphicx}
\usepackage{bm}
\usepackage{float}
\usepackage{epsfig}
\usepackage{graphicx}
\usepackage{rotating}
\usepackage{amsmath}
\usepackage{amsfonts}
\usepackage{amssymb}

\newcommand{\sqrts}{\sqrt{s}}
\newcommand{\sqrtsnn}{\sqrt{s_{_{\ensuremath{\it{NN}}}}}}
\newcommand{\pp}{$p$-$p$}
\newcommand{\ppbar}{$p$-$\bar p$}

\newcommand{\QQbar}{Q \bar Q}

\newcommand{\Lumi}{\mathcal{L}}
\newcommand{\Lunits}{cm$^{-2}$s$^{-1}$}
\newcommand{\mH}{m_{_{\rm H}}}

\def\mean#1{\ensuremath{\left<#1\right>}}

\newcommand{\MET}{\not\!\!E_{T}}
\newcommand{\ecrit}{\varepsilon_{\mbox{\tiny{\it crit}}}}

\newcommand\qhat{{\mean{\hat{q}}}}

\newcommand{\dzero}{D$\emptyset$}

\begin{document}
\title{Physics at the LHC: a short overview}

\author{David d'Enterria}
\address{ICREA \& ICC-UB, Universitat de Barcelona, 08028 Barcelona, Catalonia}
\address{CERN, PH Department, 1211 Geneva, Switzerland}


\begin{abstract}
The CERN Large Hadron Collider (LHC) started operation 
a few months ago. The machine will deliver proton-proton and nucleus-nucleus
collisions at energies as high as $\sqrts$~=~14~TeV and luminosities up to $\Lumi~\sim$~10$^{34}$~\Lunits,
never reached before. The main open scientific questions that the seven LHC
experiments -- ATLAS, CMS, ALICE, LHCb, TOTEM, LHCf and MOEDAL -- aim to solve 
in the coming years are succinctly reviewed.
\end{abstract}

\vspace{-0.5cm}
\section{Introduction}
\vspace{0.15cm}
The LHC~\cite{lhc} is the ultimate particle collider in terms of centre-of-mass (c.m.) energies 
and luminosity. The machine together with its 7 international experiments at the CERN 
laboratory in Geneva, have been built to study the frontier of our knowledge 
about the particles, interactions, and space-time structure of the Universe.  
Their design goal addresses, in particular, 
at least 6 fundamental questions that still remain unsolved today in high-energy physics:
\begin{enumerate}
\item \underline{Mass generation problem}: How do the elementary particles of the 
Standard Model (SM) acquire their bare masses ? Is it via their coupling to the Higgs boson
-- the last predicted missing piece of the SM -- or through other mechanisms ?
\item  \underline{Hierarchy / fine-tuning problem}: 
What mechanism stabilizes the Higgs boson mass up to the next known physics scale at
Planck energies ($10^{16}$ orders-of-magnitude above) ? Super-symmetry (SUSY) ? 
Higgs boson compositeness ? new space dimensions ?
\item \underline{Dark matter (DM) problem}: What weakly-interacting particle accounts for one 
fourth of the (invisible) content of the universe ? Can the lightest sparticle or other 
new massive particles (lightest technihadron or Kaluza-Klein tower, axions, sterile $\nu$'s, ...) explain DM~? 
\item \underline{Flavour problem}: Why do we observe a matter-antimatter asymmetry in the Universe~? 
Why are there so many types of matter particles and they mix the way they do~?
\item \underline{Strong interaction in non-perturbative regime}: Why quarks are confined in hadrons~?
What is the energy evolution of the total hadronic cross sections~? Can one experimentally test the 
conjectured duality between gauge and string theories (AdS/CFT)~?
\item \underline{Origin/Nature of the highest-energy cosmic-rays (CRs)}: What are the sources and
type of particles constituting CRs at energies up to 10$^{20}$ eV ?
\end{enumerate}
Of course, the solutions to these open problems need to be consistent, if not directly connected,
among each other. 
The two large multipurpose detectors, ATLAS~\cite{atlas_tdr} and CMS~\cite{cms_tdr}, have the experimental 
capabilities to address the full set of questions. The two mid-size experiments, LHCb~\cite{lhcb} and
ALICE~\cite{alice}, are mostly optimised for problems (iv) and (v) respectively, although they 
can cover a subset of the rest of research topics. 
The smallest ones, TOTEM~\cite{totem}, LHCf~\cite{lhcf} and MOEDAL~\cite{moedal}, 
aim at studying particular aspects of questions (v), (vi) and (iii) respectively.

\section{Preface: Rediscovering the Standard Model (SM)}
\vspace{0.15cm}

The SM is a renormalizable quantum-field-theory which -- unifying quantum mechanics 
and special relativity -- explains the fundamental 
interactions (except gravity) among elementary particles via a local 
SU$_{\rm C}$(3)$\times$SU$_{\rm L}$(2)$\times$U$_{\rm Y}$(1) gauge-symmetry group\footnote{The
subindices indicate the conserved {\underline c}olor and {\underline w}eak-hypercharge 
and the action on {\underline l}eft-handed fermions.}. 
The three gauge-symmetry terms give rise to the strong, weak and 
electromagnetic interactions. The particles fall into different representations of these 
groups. The SM Lagrangian (without neutrino masses) contains 19 free parameters: 
3 gauge couplings, 6 quark masses, 3 lepton masses, 3 mixing-angles, 2 CP phases, and 
2 Higgs-boson couplings, to be determined experimentally.
The SM (except the Higgs sector) has been verified to high precision in many measurements 
and the LHC will allow one to test it in a regime of energies up to 7 times higher than those
probed before. Prior to looking for new physics signals one needs to confirm that we have 
a good experimental and theoretical control of processes which have been already measured 
at lower energies.

\begin{figure}[htpb]
\begin{minipage}{8.3cm}
\includegraphics[width=8.3cm,height=10.5cm,clip]{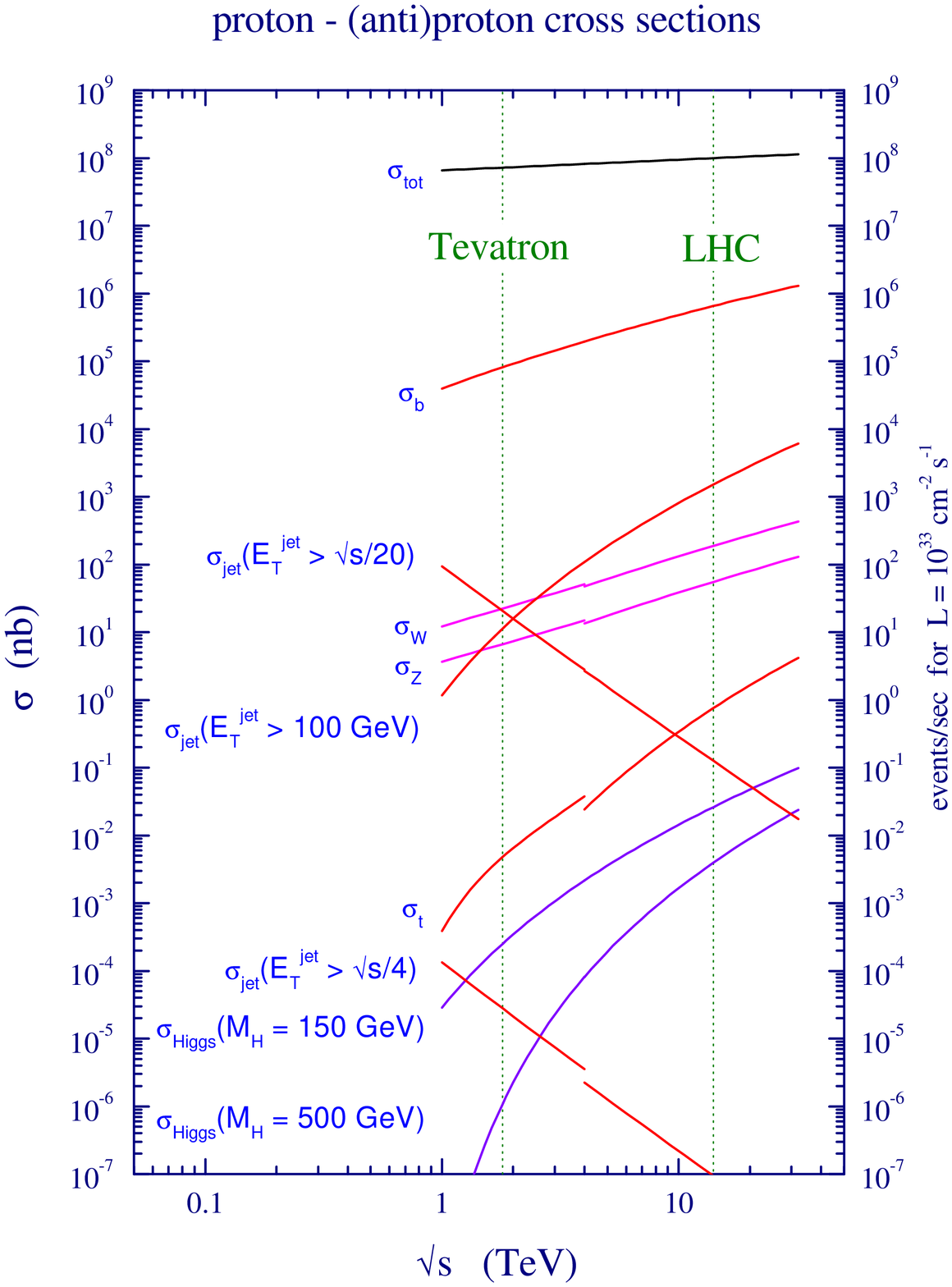}
\end{minipage}
\hspace{0.3cm}
\begin{minipage}{9.5cm}
\includegraphics[height=5.5cm,width=7.cm]{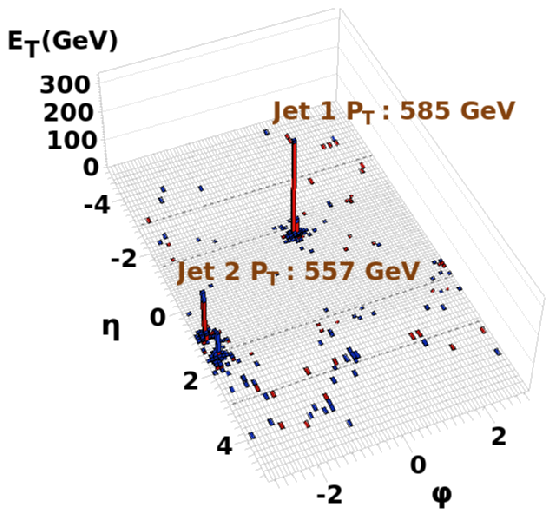}
\includegraphics[height=7.cm,angle=-90,clip]{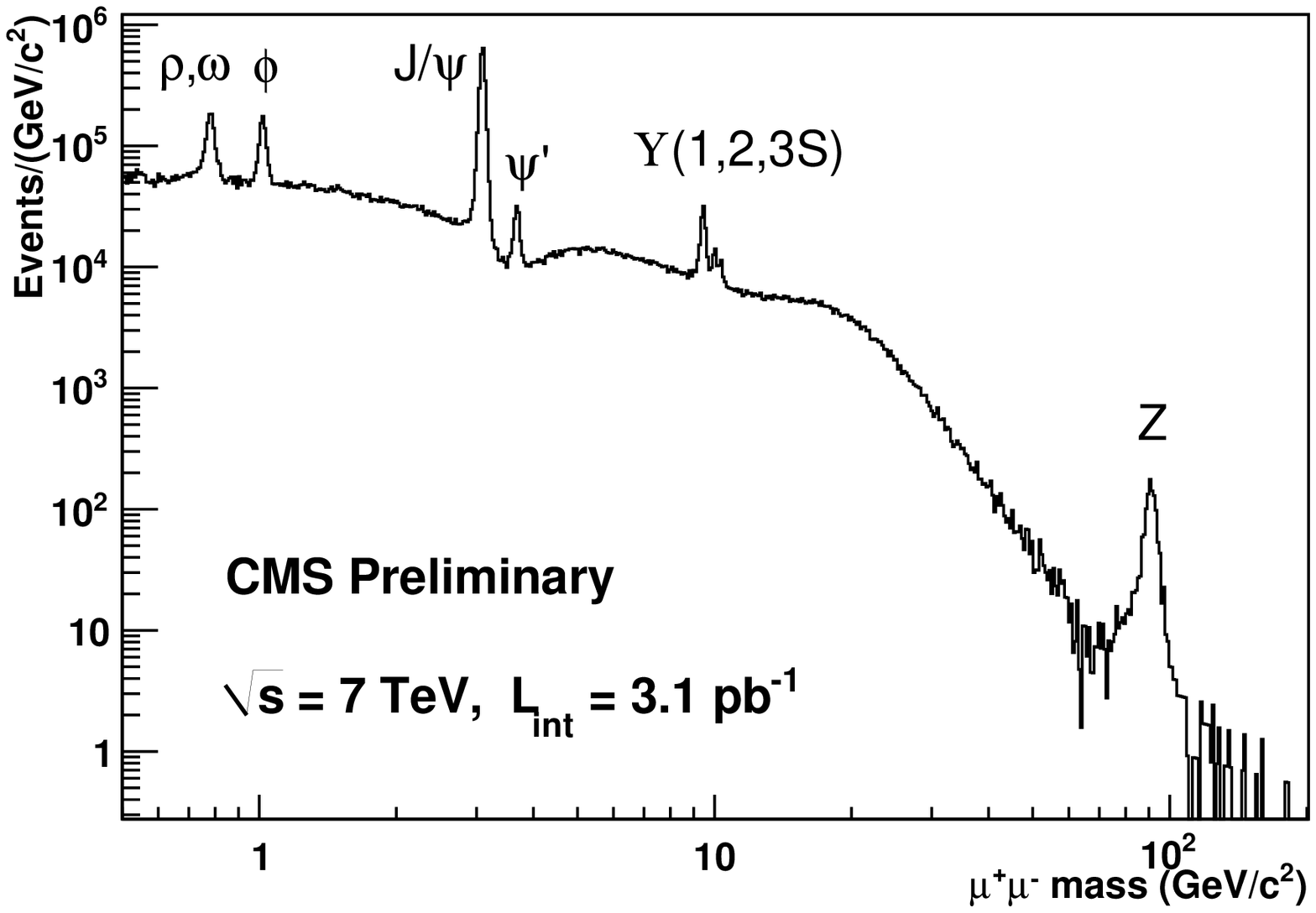}
\end{minipage}
\caption{Left: Cross sections (and rates for $\Lumi$~=~10$^{34}$~\Lunits) as a function of $\sqrts$ 
for various SM 
processes in \pp(\ppbar) collisions~\protect\cite{Campbell:2006wx}. 
Right: First measurements in \pp\ at 7~TeV: dijet event with $m_{\rm jj}$~=~2.13~TeV/c$^2$ (top),
and mass spectrum of all known dimuon resonances (bottom)~\protect\cite{cms_ichep10}.}
\label{fig:SM}
\end{figure}

The processes with the largest cross sections in hadronic collisions are mediated 
by the strong force (upper curves in Fig.~\ref{fig:SM}, left) and thus 
the first LHC measurements in \pp\ at $\sqrts$~=~7~TeV are linked to QCD observables 
such as those shown in Fig.~\ref{fig:SM} right. Perturbative QCD calculations 
agree well with preliminary data such as high-$p_T$ hadrons, jets, and heavy-flavours 
measured by ATLAS~\cite{atlas_ichep10}, CMS~\cite{cms_ichep10}, ALICE~\cite{alice_ichep10} 
and LHCb~\cite{lhcb_ichep10}. Next in importance in terms of cross sections are the electroweak
processes: the first W and Z bosons (highest-mass dimuon peak in Fig.~\ref{fig:SM} right)
have been detected with the first hundred nb$^{-1}$ integrated by CMS and ATLAS. 

\section{Problem I: Mass generation problem -- the Higgs boson }
\vspace{0.15cm}

The LHC together with the ATLAS and CMS experiments have been designed primarily to be able to 
solve the last missing element of the SM: the mechanism of generation of the elementary 
particle masses. Indeed, the electroweak (EWK) sector of the theory suffers from two problems: 
\begin{itemize}
\item The SU$_{\rm L}$(2)$\times$U$_{\rm Y}$(1) symmetry imposes zero-masses for the gauge bosons 
and fermions in striking contradiction, in particular, with the large observed masses for W and Z.
\item Without a mechanism to generate the vector-boson masses, the longitudinal WW scattering 
cross sections grow quadratically with energy and break unitarity at energies $\mathcal{O}$(1~TeV).
\end{itemize}
A simple and elegant solution to these problems, was proposed by 
Englert-Brout-Higgs-Guralnik-Hagen-Kibble~\cite{EBHGHK} who demonstrated that the W and Z
bosons can acquire a mass while preserving the SU$_{\rm L}$(2)$\times$U$_{\rm Y}$(1) 
symmetry of the EWK Lagrangian, if the vacuum is filled with a field with a non-zero expectation 
value which 
couples to the (massless) vector-bosons. The electroweak 
symmetry is preserved in the Lagrangian but it is broken spontaneously by the ground-state 
actually realized in nature. Such a mechanism involves adding new 
terms, an extra scalar-doublet and associated couplings, to the SM Lagrangian.
Three of the four degrees of freedom of this extra field mix with the W,Z bosons to provide
them with mass, while the other one becomes the Higgs boson, a new scalar particle.
In a similar manner, Higgs bosons ``lurking virtually'' in the vacuum drag on all fermions 
to give them mass. The quadratic rise of the WW,ZZ cross sections is thus
damped by new diagrams involving Higgs boson exchange.




\begin{figure}[htbp]
\begin{minipage}{7.5cm}
\includegraphics[height=7.5cm,width=6.cm,angle=-90,clip]{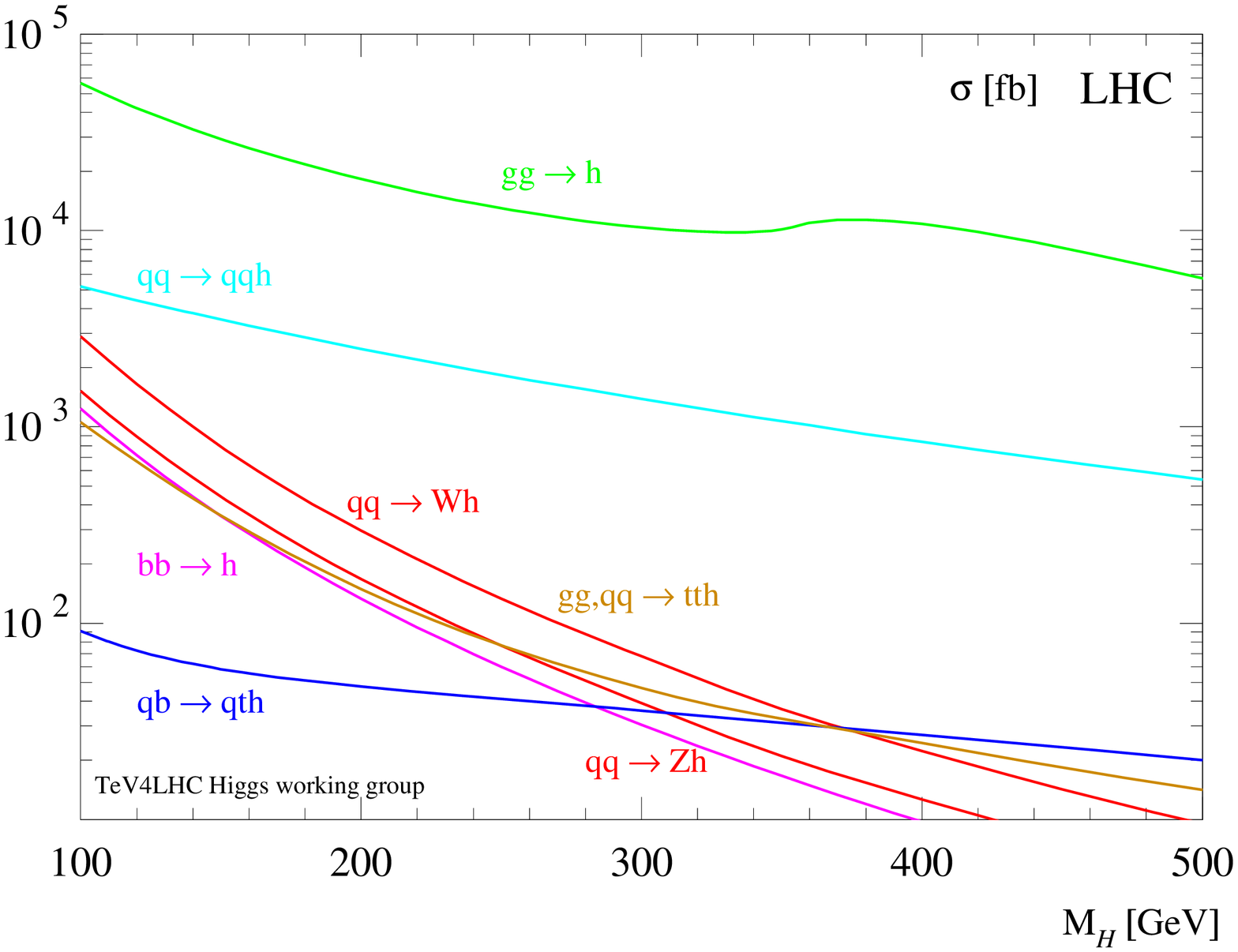}
\end{minipage}
\begin{minipage}{8.5cm}
\includegraphics[width=8.5cm,height=6.cm]{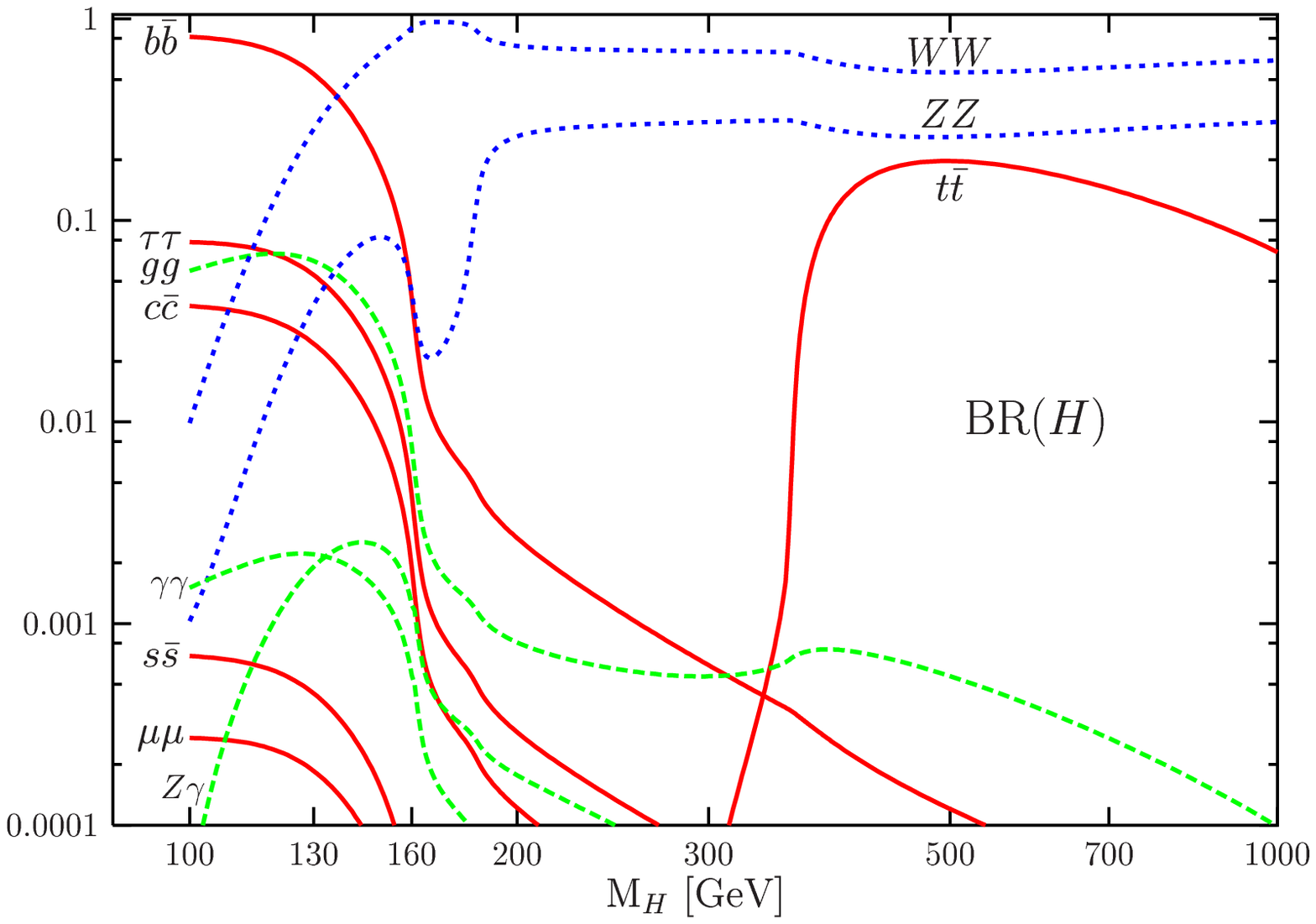}
\end{minipage}
\caption{SM Higgs boson:\,LHC production channels (left)\,\cite{Aglietti:2006ne} and 
branching ratios (right)\,\cite{Djouadi:2005gi}.}
\label{fig:higgs}
\end{figure}

The mass of the Higgs boson is a free parameter of the SM. Direct searches at 
LEP~\cite{Barate:2003sz} and Tevatron~\cite{ichep2010} have 
constrained its value at 95\% confidence level (CL) to the range 114.4~--~158~GeV/c$^2$
and $\mH >$~175~GeV/c$^2$, whereas indirect constraints from global fits to precision 
EWK data (accounting for virtual Higgs contributions e.g. to the W and top-quark
masses) exclude $\mH >$~185~GeV/c$^2$ at 95\% CL~\cite{Collaboration:2008ub}.
At the LHC, several Higgs production processes and decay channels are accessible 
(Fig.~\ref{fig:higgs}). 
Depending on its mass, different production/decay modes can be exploited for a 
5$\sigma$-discovery integrating a few tens of fb$^{-1}$ in \pp\ 
at 14-TeV~\cite{atlas_tdr,cms_tdr}:
\begin{enumerate}
\item For $\mH<$~135 GeV/c$^2$, the preferred channels involve $\gamma\,\gamma$ and 
$\tau\,\tau$ decays, in particular in vector-boson-fusion production processes accompanied by forward-backward jets.
\item For $\mH>$~135 GeV/c$^2$, the $H\to WW^{(*)},ZZ^{(*)}$ modes, characterized by four 
(two) high-$p_T$ electrons or (and) muons in the final-state, provide a relatively clean signal.
\end{enumerate}
The determination of the Higgs couplings to all SM particles and its quantum numbers
will likely require a more precise $e^+e^-$ collider or resort to other more
suppressed production channels~\cite{d'Enterria:2009wn}.

\section{Problem II: Hierarchy/fine-tuning -- New symmetries at high energies ?}
\vspace{0.15cm}

The hierarchy problem of the SM is related to the ``uncontrollable'' running of the Higgs boson 
mass with energy. Indeed, 
(i) the $\mH$ value is not predicted by the theory, nor protected by any internal SM symmetry;
and (ii) being a scalar field, the loop radiative corrections make $\mH$ increase 
{\it quadratically}\footnote{In contrast, the radiative corrections for the fermion masses 
are only logarithmic with energy.} up to next physics scale known today, where gravity becomes as strong as the 
gauge interactions. Thus, if $\mH$ is imposed from a symmetry at the Planck scale, then its value has to be 
fine-tuned to 1 part in $m_{_{\rm EW}}/m_{_{\rm Planck}}\approx G_{\rm F}^{-1/2}/G_{\rm N}^{-1/2}\approx 10^{16}$ !


\begin{figure}[htbp]
\includegraphics[width=3.8cm,height=2.45cm]{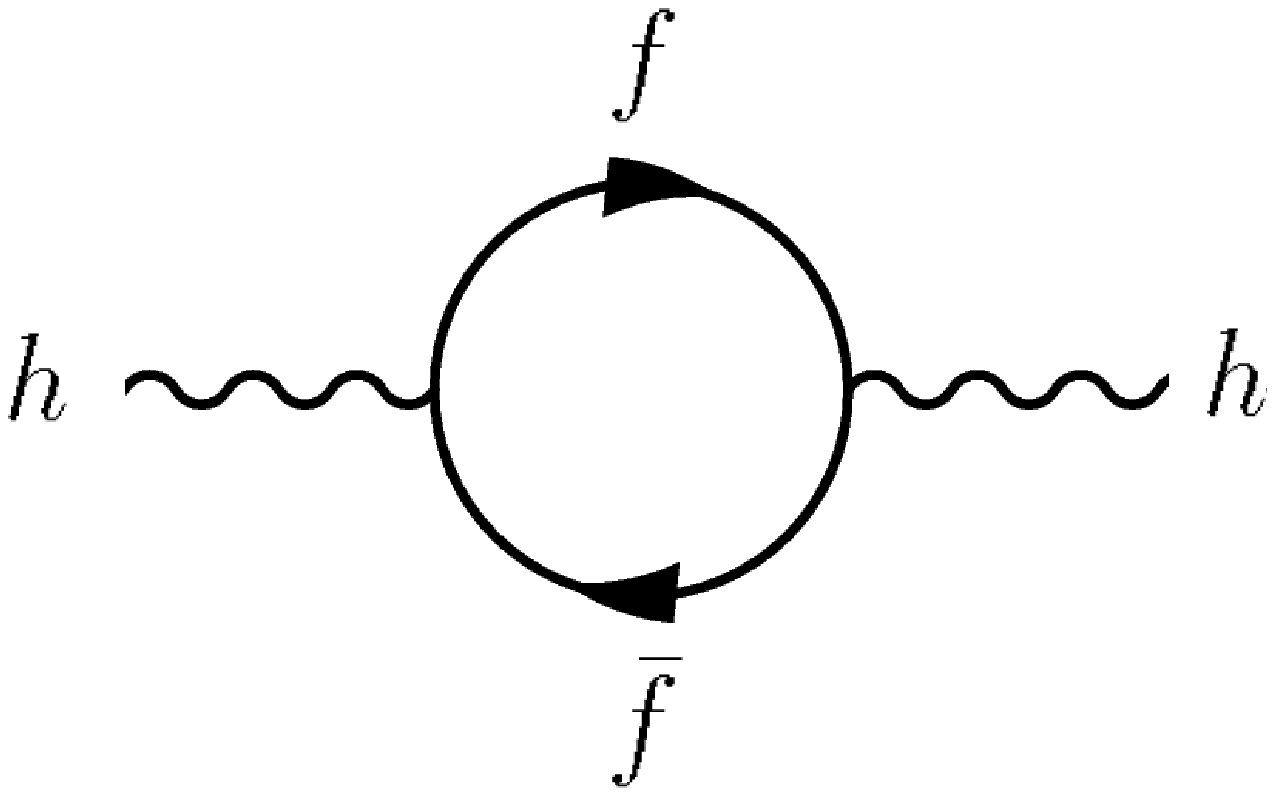}\hspace{0.3cm}
\includegraphics[width=4.0cm,height=2.5cm]{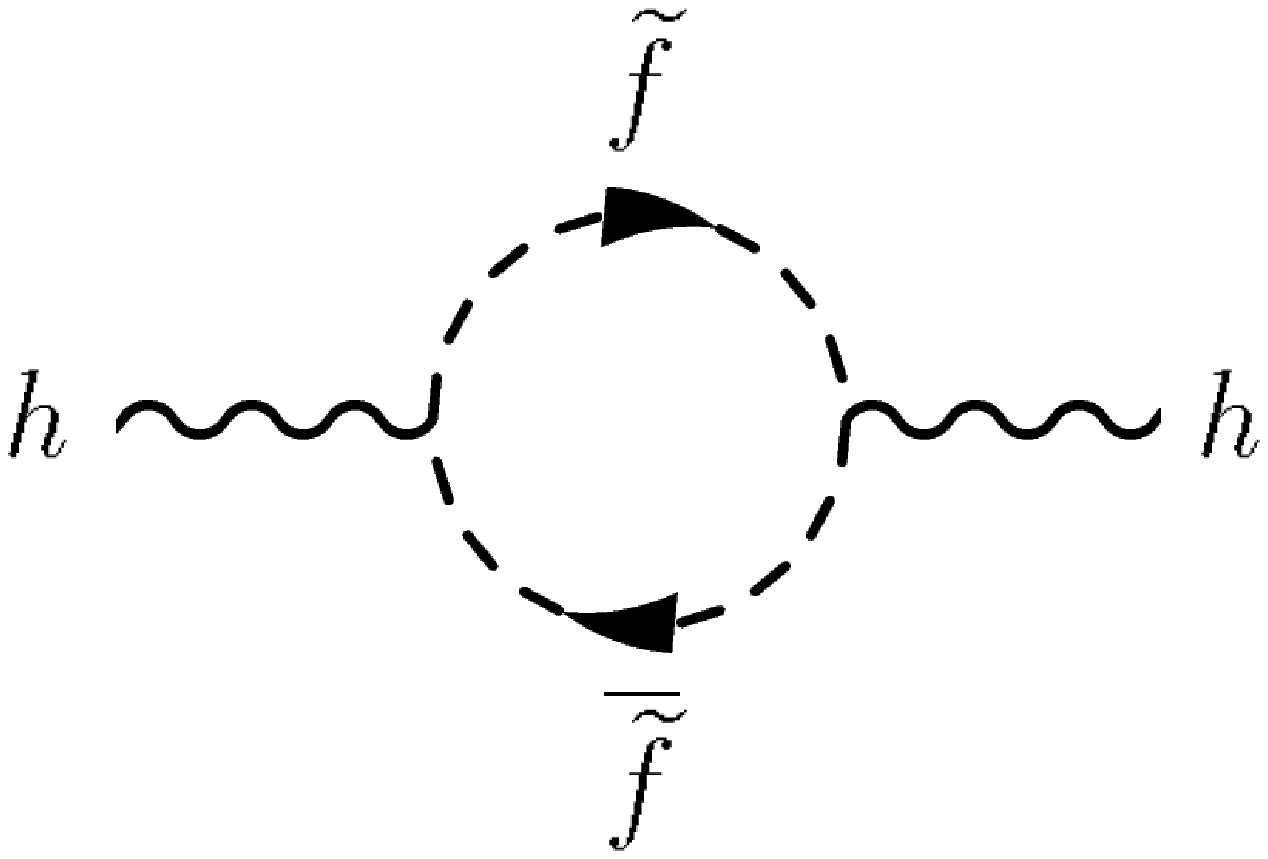}\hspace{0.5cm}
\includegraphics[height=3.2cm]{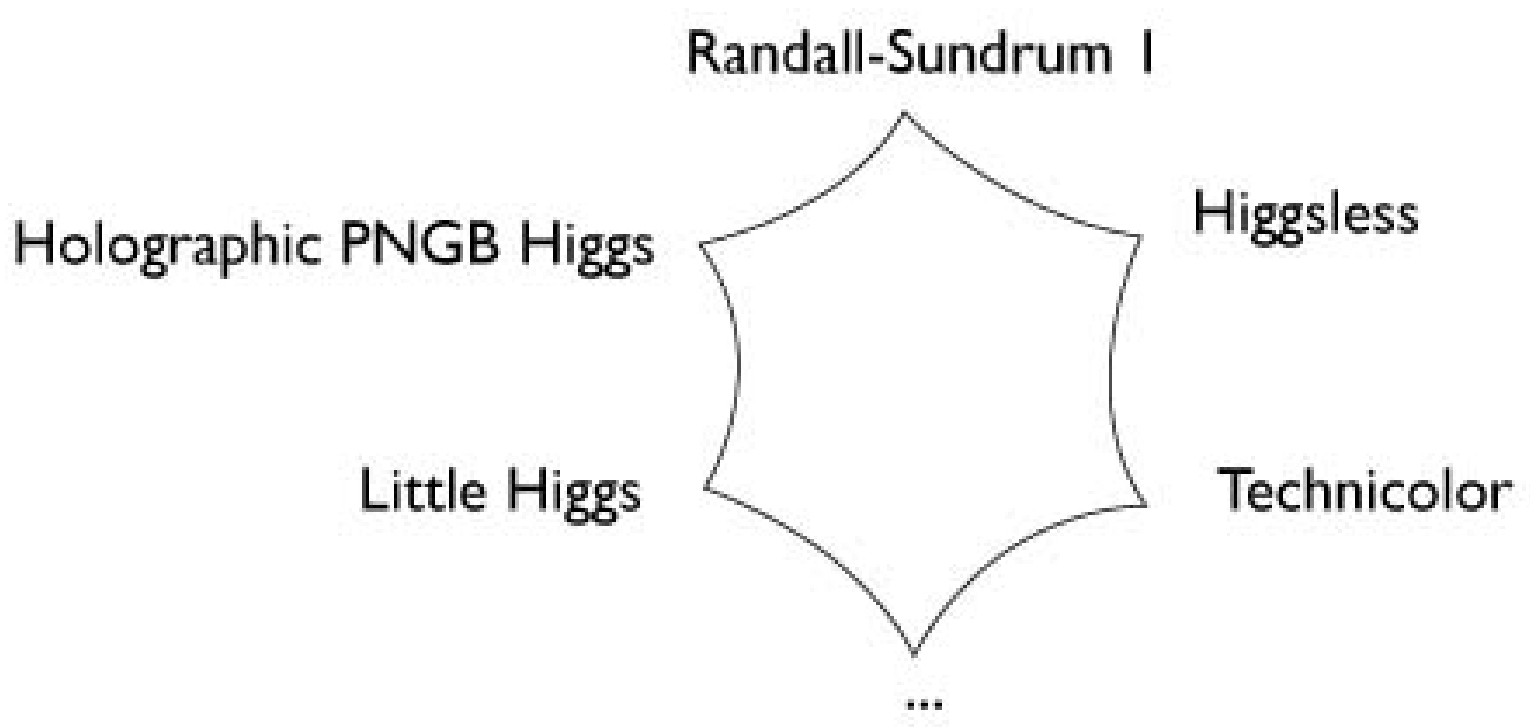}
\caption{Left: Virtual contributions to the Higgs mass from SM fermions and SUSY sfermions.
Right: Alternative (non-SUSY) models beyond the SM~\cite{Cheng:2007bu}.}
\label{fig:bsm}
\end{figure}

There are three general theoretical solutions to the hierarchy problem. All of them imply 
physics beyond the Standard Model (BSM), and the existence of new symmetries and associated 
new particles at the TeV scale:
\begin{enumerate}
\item SUSY: The existence of new SUSY partners, differing by 1/2 unit of spin, for every SM particle
provides a simple way to stabilize the Higgs potential, as their amplitude in the quantum 
corrections come with opposite sign and cancel the SM contributions (Fig.~\ref{fig:bsm}, left).
\item Non-Standard ``Higgs'' models: The W and Z masses can also be generated {\it dynamically} 
via a Goldstone boson corresponding to a spontaneously broken global symmetry of a new 
strongly-interacting (QCD-like) sector at some higher mass scale $\Lambda$, as done e.g. 
in Technicolour~\cite{technicolor} or little-Higgs models. In the previous case, the Higgs boson is not an
elementary scalar but a condensate of techni-fermions. 
In the latter, the Higgs is a pseudo-Goldstone boson and its mass is protected from acquiring quadratically-divergent 
loop corrections by the contributions of new particles (heavy-top, W', Z', ...).
\item Low-energy quantum gravity: The huge hierarchy between the EWK and Planck scales 
can be solved if one considers that 
the apparent relative weakness of gravitation is not real but only due to the fact that it expands over
extra (hidden) spatial dimensions, whereas the other interactions are confined to 
our visible 3-D space. Theories with extra dimensions (flat as in ADD~\cite{ADD} or warped as in RS~\cite{RS} models) 
predict also new particles at the TeV-scale (Kaluza-Klein towers, radion, mini-blackholes, ...).
\end{enumerate}

\section{Problem III: Dark matter -- New heavy particles ?}
\vspace{0.15cm}

The existence of dark matter (DM), accounting for one-fourth of the content of the Universe has been
confirmed by several experimental evidences such as
(i) the fact that the galactic rotation speed curves do not follow the expected Copernican fall-off 
with distance from the center, (ii) the structure of the power spectrum of the temperature fluctuations 
of the cosmic microwave background, (iii) the separation in the distribution of matter observed by 
gravitational-lensing and by radiation in the collision of two clusters of galaxies. 
The current DM signatures favour a Weakly-Interacting Massive Particle (WIMP) which
is sensitive only to the weak-interaction and gravitation, and which is a stable heavy 
relic from the early Universe.
All BSM theories mentioned in the previous section naturally contain DM candidates such as:
\begin{enumerate}
\item the lightest SUSY Particle (LSP), such as neutralinos or gravitinos;
\item the lightest technibaryon in technicolor models;
\item the lightest Kaluza-Klein tower (resonances arising from quantized waves in the extra dimension), 
gravitons from adjacent branes, or radions; in ADD or RS approaches.
\end{enumerate}
In addition, axions or heavy right-handed (sterile) neutrinos, among others have also been
proposed as DM particle candidates.
\begin{figure}[htbp]
\begin{minipage}{7.5cm}
\includegraphics[width=7.5cm,height=5.2cm]{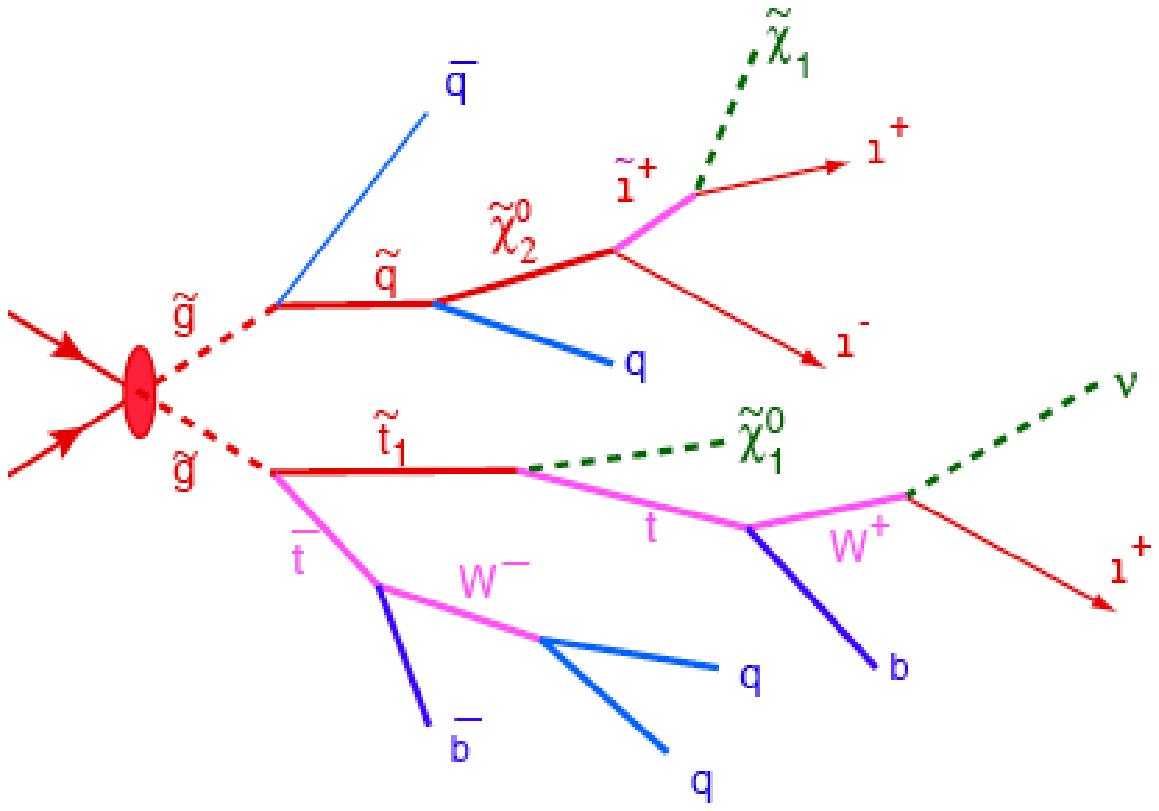}
\end{minipage}
\begin{minipage}{8.7cm}
\includegraphics[width=8.7cm,height=5.8cm]{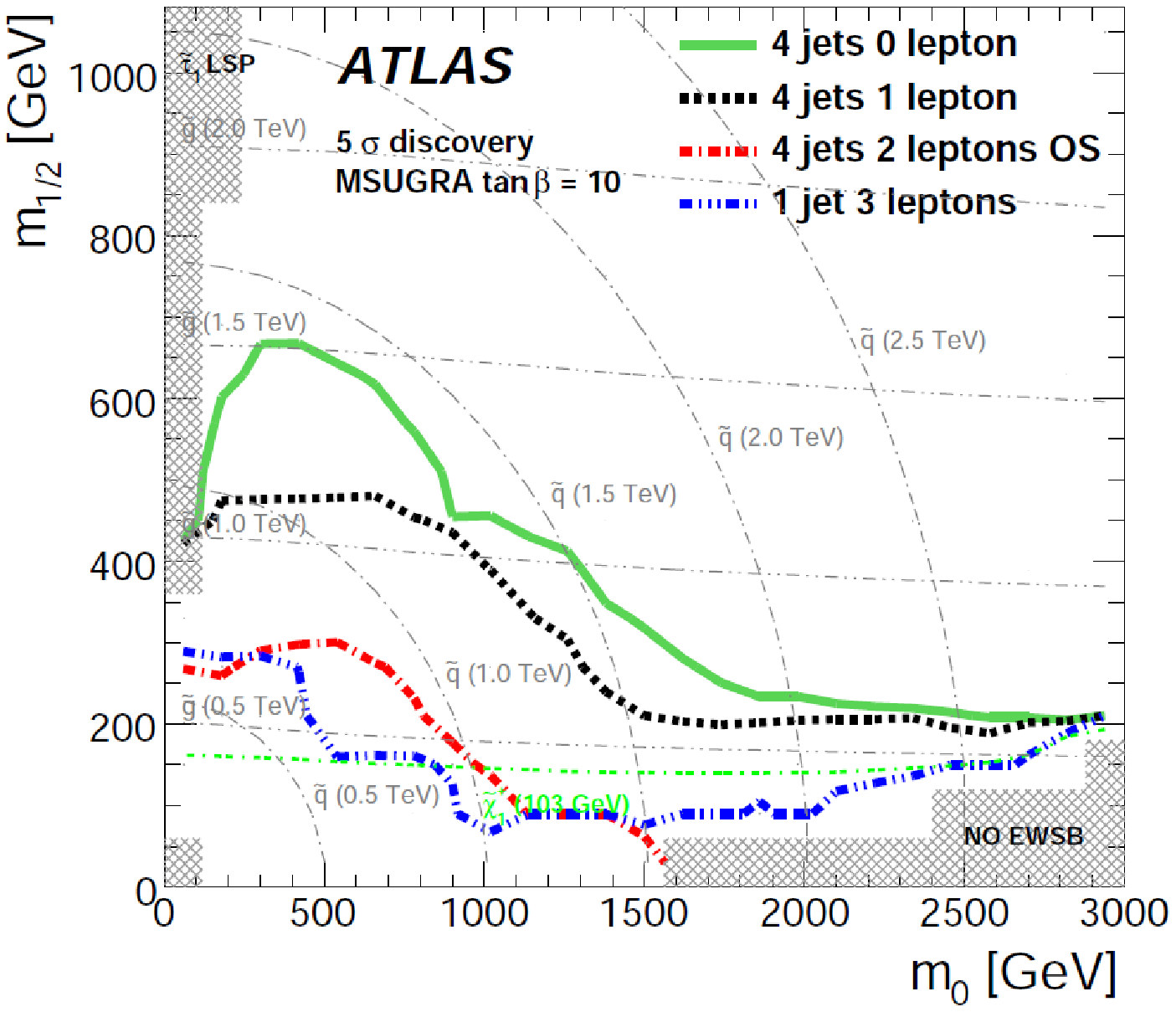}
\end{minipage}
\caption{Left: Cascade decay of a pair of gluinos in a \pp\ collision.
Right: mSUGRA discovery contours in jets+leptons+$\MET$ channels 
(\pp\ 14 TeV, 1 fb$^{-1}$) as a function of $m_0$ and $m_{1/2}$~\cite{atlas_tdr}.}
\label{fig:susy}
\end{figure}

By its own characteristics, any heavy WIMP produced at the LHC will appear as (large) 
missing transverse energy in a \pp\ event. 
In the SUSY case\footnote{The LSP is stable, as it cannot decay anymore, if SUSY
conserves R-parity 
(i.e. sparticles are produced in pairs and decay into another SUSY particle plus any number of normal particles).} 
the LSP -- e.g. the neutralino (a neutral-charge mixture of superpartners of the SM gauge bosons) 
or the gravitino -- is a paradigmatic stable heavy candidate for DM. LSPs are produced in events
whose experimental event topologies (Fig.~\ref{fig:susy}, left) are characterized by:
\begin{itemize}
\item missing transverse energy ($\MET$) from the invisible WIMP at the end of a decay cascade,
\item multi-jets: from intermediate sparticle decays into heavy SM particles (top, W, Z),
\item same-sign leptons/multi-photons: from decay of the intermediate sparticles.
\end{itemize}
An example of the discovery potential of a particular class of SUSY models 
(mSUGRA, $\tan\beta = 10$) based on such experimental signatures, is shown in 
Fig.~\ref{fig:susy} (right) as a function of two of the free parameters of the theory
(the common scalar $m_0$ and fermion $m_{1/2}$ masses).\\

In addition, in many BSM approaches the next-to-lightest new particle is metastable 
(long-lived) and has a mass/charge ratio which makes it a highly ionizing particle 
when interacting with normal matter\footnote{Typical SUSY highly-ionizing particles 
are NLSPs such as staus as well as R-hadrons (gluino-parton bound-states).}. 
The MOEDAL experiment~\cite{moedal} (sharing the interaction point with LHCb) 
aims at measuring such highly-ionizing tracks in large plastic detectors.

\section{Problem IV: Matter-antimatter asymmetry -- new virtual particles ?}
\vspace{0.15cm}

\begin{figure}[htbp]
\begin{minipage}{7.10cm}
\includegraphics[width=7.1cm]{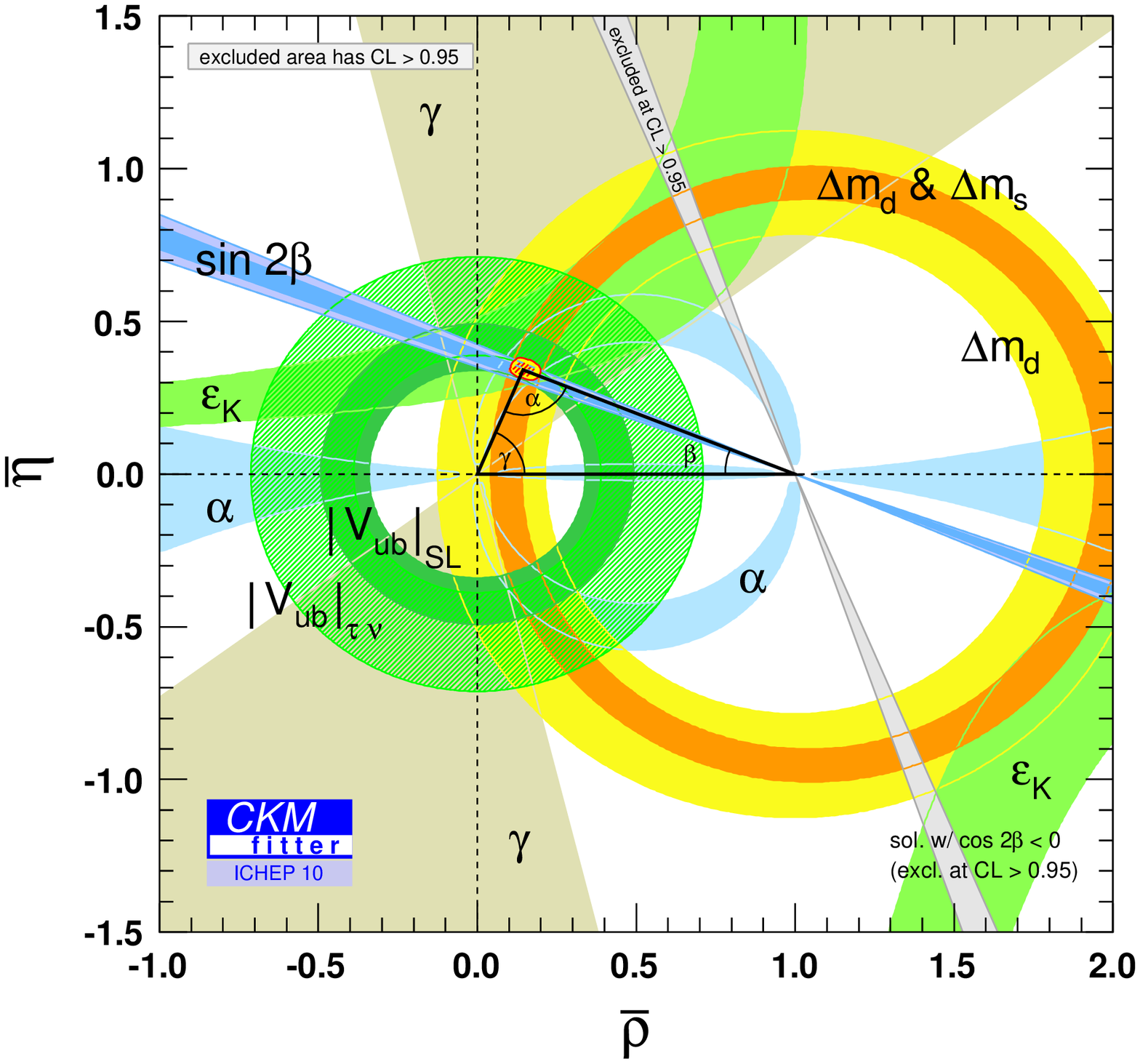}
\end{minipage}
\hspace{0.3cm}
\begin{minipage}{8.cm}
\centering
\includegraphics[width=5.8cm,height=3.3cm]{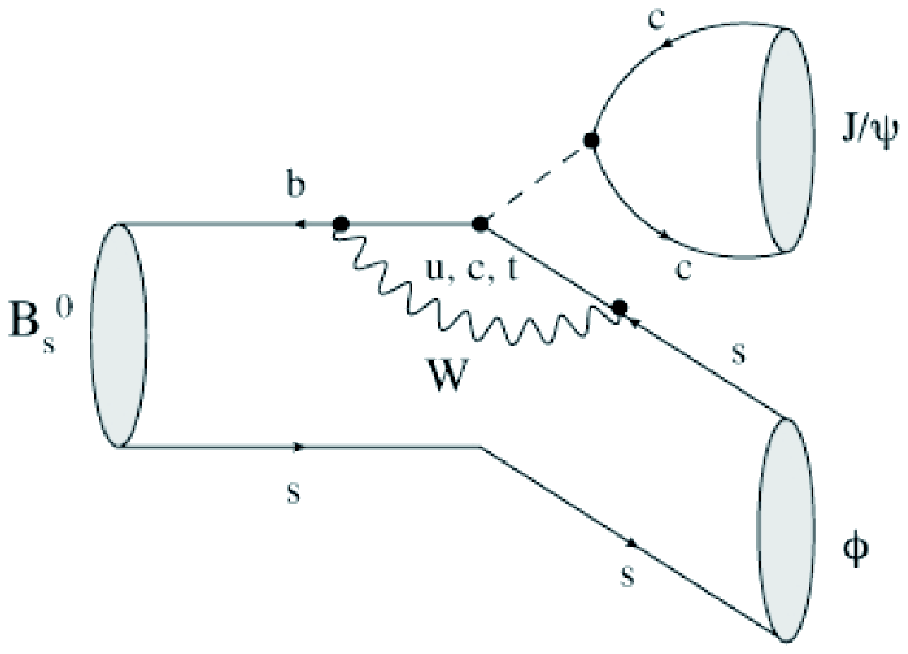}
\includegraphics[width=7.8cm,height=2.4cm]{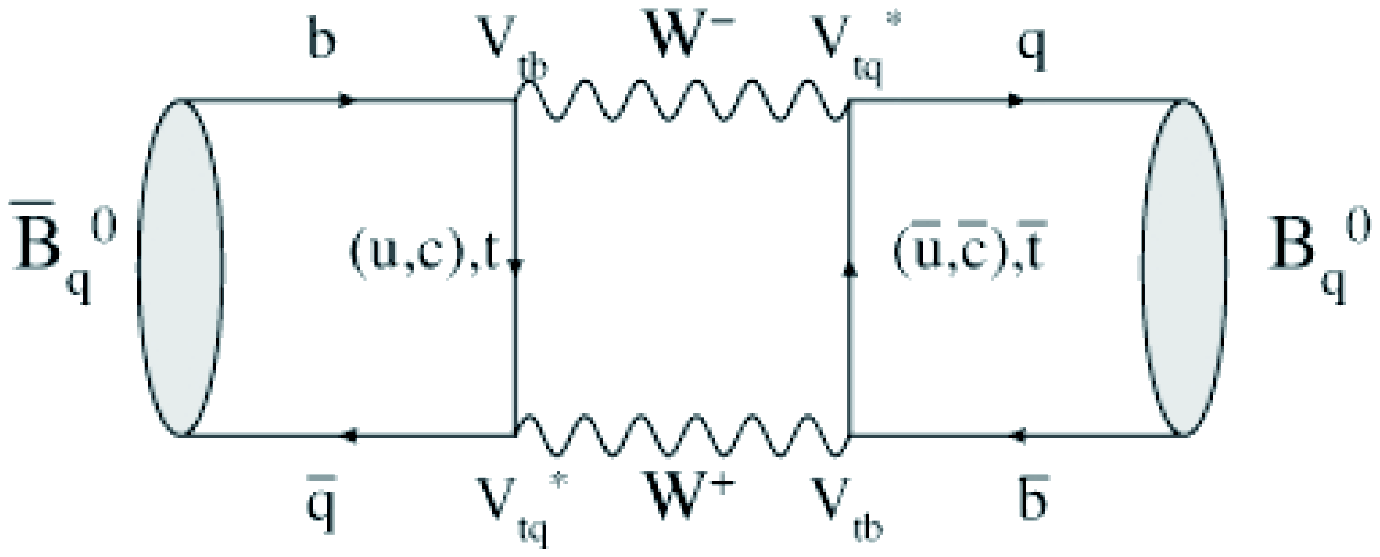}
\end{minipage}
\caption{Left: Unitarity triangle fit~\cite{ckmfitter}. 
Right: Examples of penguin (top) and box (bottom) diagrams for B-meson decay and oscillations in the SM.
\label{fig:CP}}
\end{figure}

The process of changing a quark flavour into another is controlled by 
the (charged-current) weak interaction. The observation that kaons and 
B-mesons can transform into their anti-particles with different 
probabilities in each direction and that they can decay into a state with 
different Charge-Parity (CP) quantum numbers, implies that 
the weak force is not invariant under C- (particles and antiparticles exchange)
and P- (changing a particle by its mirror image) transformations. In the SM,
weak interactions involving W exchange act exclusively on left-(right-)handed 
(anti)particles, and the coupling among quarks is done in terms of mixed-flavour 
objects which do not correspond to the quark mass-eigenstates (to which QCD and QED couple) 
but to a superposition of them. The relationship between the mass- and weak- eigenstates 
is implemented via the Cabibbo-Kobayashi-Maskawa (CKM) matrix $V_{ij}$, which describes 
how quarks mix/decay among each others. CP violation is then incorporated via a CKM 
matrix complex phase, constrained by the condition $V_{ud}V_{ub} + V_{cd}V_{cb} + V_{td}V_{tb} = 0$, 
which is often displayed as a unitarity triangle with angles $\alpha=f(V_{td,tb},V_{ud,ub})$, 
$\beta=f(V_{cd,cb},V_{td,tb})$ and $\gamma=f(V_{ud,ub},V_{cd,cb})$ obtained 
from kaon and B-meson measurements (Fig.~\ref{fig:CP}, left).

The currently known differences between particles and antiparticles (i.e. 
the amount CP-violation in the SM) are however way too small 
to explain the  matter-antimatter imbalance observed today,
$(n_{_{\rm B}}-n_{_{\rm \bar{B}}})/n_{_{\gamma}}\approx 10^{-9}$,
and new particles/CP-phases are needed to explain how baryon dominance
appeared in the Universe (baryogenesis). CP-violation studies at the LHC involve 
indirect searches of new 
virtual particles contributing to higher-order (Penguin or box) loops 
in flavour-changing charged current processes (Fig.~\ref{fig:CP}, right).
The LHCb experiment~\cite{lhcb} focuses 
on measurements in the bottom sector (e.g. $b\to s$ transitions) 
which are less constrained by current data 
and can access higher energy scales than the kaon system, such as: 
\begin{enumerate}
\item detailed B-meson studies of rare decay rates, branching ratios,
decays asymmetries, oscillation frequencies, and lifetimes;
\item over-constraints of the unitarity triangle via improved precision of 
CKM angles and sides, and cross-checks of the same quantity via various measurements.
\end{enumerate}
In all cases, any possibly found inconsistency (in the phases of the couplings, their
absolute values, and/or their Lorentz structure) can be a sign of new physics.

\section{Problem V: Strongly-interacting matter -- QGP, CGC, AdS/CFT, $\sigma_{\rm tot}$}
\vspace{0.15cm}

\begin{figure}[htbp]
\begin{minipage}{6.0cm}
\includegraphics[width=6.cm,height=4.8cm]{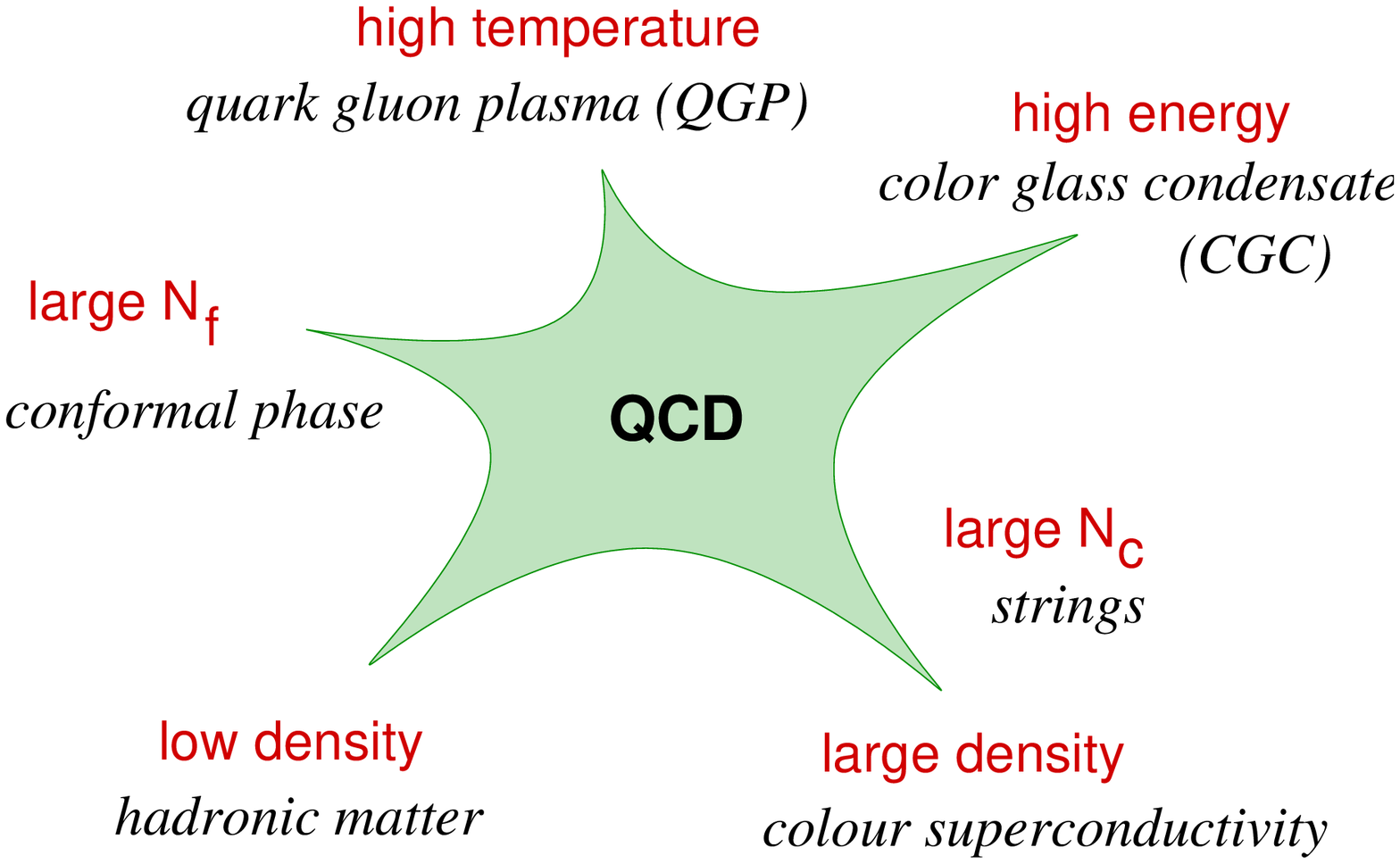}
\end{minipage}
\hspace{0.1cm}
\begin{minipage}{10.cm}
\resizebox{!}{2.3cm}{
\begin{tabular}{llll}\\\hline
Heavy-ion    & Experimental & QCD medium  & Theoretical \\ 
measurement  & observable & property   & tool \\\hline
$dN_{\rm h}/d\eta$ (soft) & multiplicity & gluon $xG(x,Q^2)$ & CGC\\
$dN_{\rm h}/dp_{_T}$ (soft) & radial flow & Eq. of State $p(T)$ & Hydro+lattQCD\\
$dN_{\rm h}/d\phi$ (soft) & elliptic flow & shear viscosity $\eta$ & Hydro+AdS/CFT\\
$dN_{\gamma}/dp_{_T}$ (semihard) & thermal $\gamma$ & temperature $T_{\rm crit}$ & Hydro+lattQCD\\
$\QQbar$ (hard) & color screening & energy density $\ecrit$ & Hydro+lattQCD\\
$dN_{\rm h,jet}/dp_{_T}$ (hard) & jet quenching & transport coeff. $\qhat$ & Hydro+pQCD\\[-2ex] 
 &  &  & \\\hline\hline
\end{tabular}
}
\end{minipage}
\caption{Left: Many-body QCD at various limits~\cite{schaefer05}. Right: Experimental and theoretical 
tools in high-energy A-A collisions ($dN_{\rm h}$ stands for differential hadron yield distributions).}
\label{fig:QCD}
\end{figure}

The strong interaction among quarks and gluons is described by QCD, a 
quantum field theory with a very rich dynamical content 
including asymptotic freedom, infrared slavery, (approximate) 
chiral symmetry, non trivial vacuum topology (instantons), strong CP problem, U$_{\rm A}(1)$ 
axial-vector anomaly, ... All these properties translate into a diverse many-body
phenomenology at various limits (Fig.~\ref{fig:QCD}, left).
Interestingly, QCD is the only SM sector whose full {\it collective} behaviour -- phase 
diagram, (deconfinement and chiral) phase transitions, thermalization of fundamental fields -- 
is accessible to scrutiny in the laboratory, via high-energy heavy-ion collisions. The study 
of strongly-interacting matter in extreme conditions of temperature, density and small parton 
momentum fraction (low-$x$) can be carried out by measuring different observables in 
nucleus-nucleus (A-A) collisions 
which are sensitive to the underlying QCD medium properties
(Fig.~\ref{fig:QCD}, right)~\cite{d'Enterria:2006su}. 

The collisions of lead nuclei at the LHC ($\sqrtsnn$~=~5.5~TeV, 30 times higher than
those attained previously) will allow one to study quark-gluon matter at unprecedented values 
of energy density ($\varepsilon_{_{\rm Bjorken}}\approx$~10~GeV/fm$^3$ at times $\tau_0$~=~1~fm/c)
using pQCD probes produced with cross sections 10 to 10$^4$ higher than at the Relativistic
Heavy-Ion Collider. The fractional momenta of the colliding partons will be as low as 
$x\approx p_T/\sqrtsnn \,exp(-\eta)$~=~${\cal O}(10^{-5}$),
where gluon saturation effects, as described in the Color-Glass-Condensate (CGC) 
approach~\cite{Gelis:2010nm}, are expected to dominate the parton dynamics.
A part from the standard searches of the formation of a deconfined and chirally-symmetric
Quark-Gluon-Plasma (QGP) via quarkonia suppression or jet quenching, one of the important 
measurements will be that of the azimuthal anisotropies of bulk hadron production with 
respect to the reaction plane. Such ``explosive'' anisotropies, known under the name of ``elliptic flow'', 
have been found to be very sensitive to the viscosity/entropy ratio of the produced medium, 
according to advanced relativistic fluid-dynamics calculations~\cite{Luzum:2008cw} 
including the QCD equation-of-state computed in the lattice. The field of heavy-ion physics 
has attracted lots of theoretical interest as testbed for the application of the
Anti-de-Sitter/Conformal-Field-Theory (AdS/CFT) duality between weakly-coupled 
gravity and strongly-coupled gauge theories~\cite{adscft}. Applications of such a
formalism have led to the determination of transport properties such as the 
viscosity~\cite{kovtun04} of Supersymmetric-Yang-Mills (SYM) plasmas, from simpler 
black-hole thermodynamics calculations. The large differences between SYM theory 
and QCD (extra SUSY degrees of freedom, no running-coupling, no confinement, ...) 
seem to ``wash out'' at finite temperature~\cite{Kajantie:2006hv}.\\


The measurements of the total and elastic \pp\ cross sections are also part of the LHC 
physics programme, providing a valuable test of fundamental quantum mechanics 
relations such as the Froissart bound $\sigma_{tot}<${\small{\it Const}} $\ln^2s$, 
the optical theorem $\sigma_{tot}\sim $Im$f_{el}(t=0)$, and dispersion relations 
Re$f_{el}(t=0)\sim$Im$f_{el}(t=0)$~\cite{Bourrely:2005qh}. 
The current extrapolations of the total \pp\ cross section at the LHC ($\sigma_{tot}$ = 90 -- 140~mb),
of which the elastic contribution accounts for about one fourth, suffer from large uncertainties due 
to disagreeing measurements at the Tevatron and uncertain extractions from cosmic-ray collisions with 
air nuclei 
(Fig.~\ref{fig:sigma_CRs}, left).
The main goal of the TOTEM experiment~\cite{totem} is to obtain a measurement of the total and elastic 
\pp\ cross sections, with an uncertainty of about 1\%, over a large range of 4-momentum transfers 
from $-t\approx 2 \cdot 10^{-3}\,$GeV$^{2}$ to 8\,GeV$^{2}$. TOTEM Roman-pots detectors will measure 
the elastically scattered protons inside the LHC tunnel area adjacent to the CMS collision point.

\section{Problem VI: Origin and nature of the highest-energy cosmic-rays}
\vspace{0.15cm}

\begin{figure}[htb]
\includegraphics[width=7.8cm,height=5.7cm]{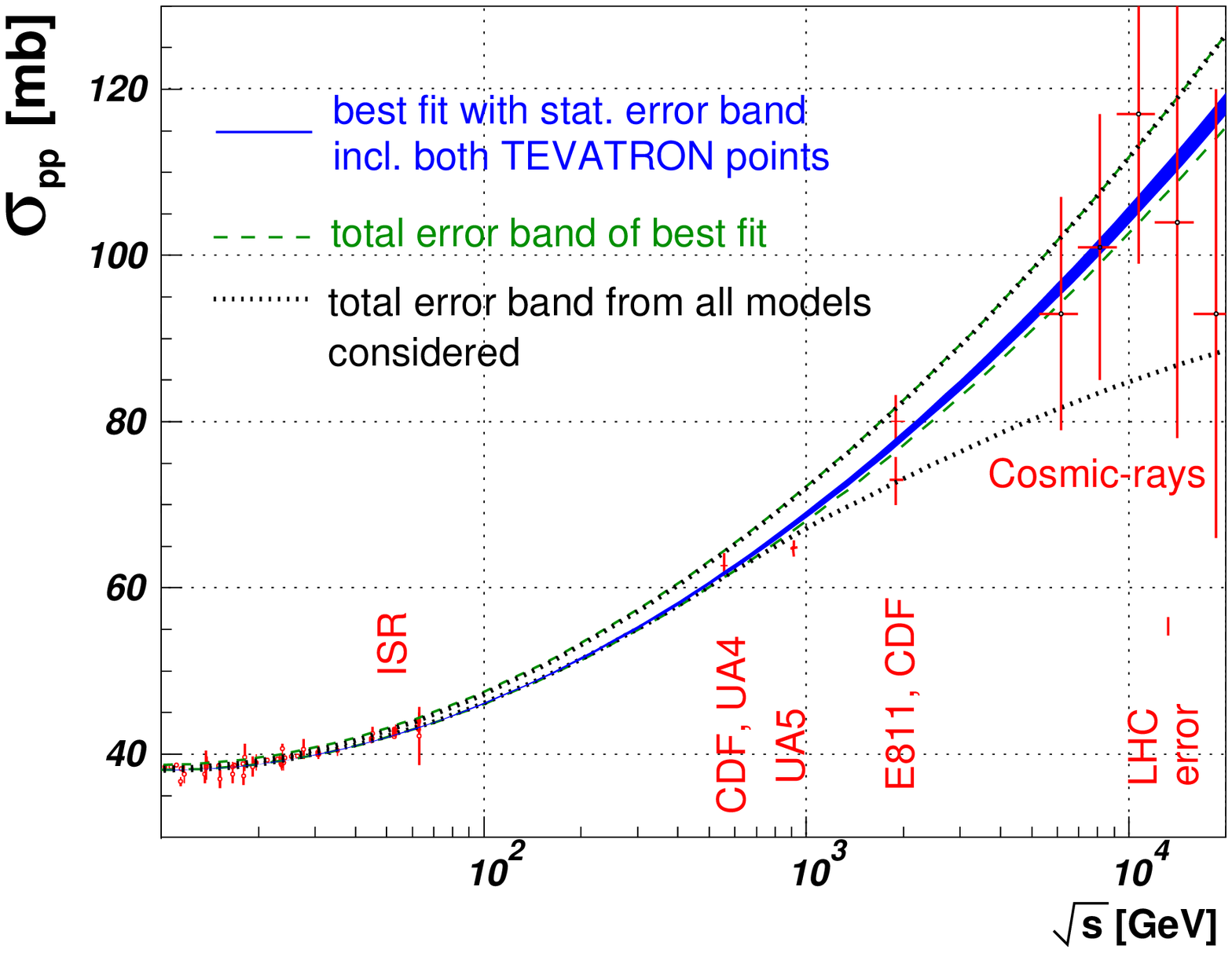}
\includegraphics[width=8.5cm]{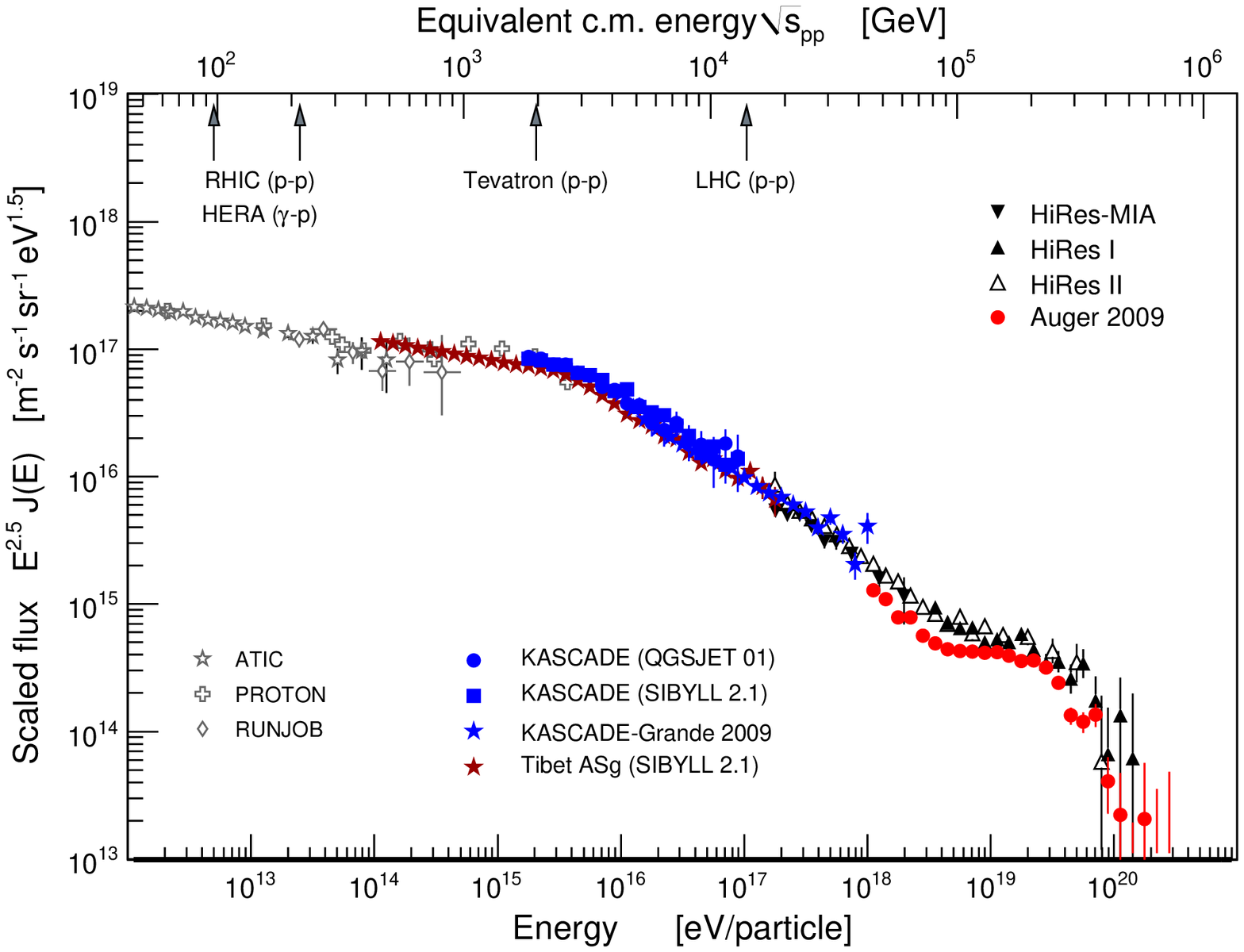}
\caption{Left: COMPETE fits~\protect\cite{compete} of $\sigma_{tot}$ as a function of $\sqrts$. 
Right: Measured (scaled) cosmic ray flux as a function of $E_{_{\rm lab}}$ (the equivalent $\sqrts$ 
of different colliders is also shown)~\protect\cite{Bluemer:2009zf}.}
\label{fig:sigma_CRs}
\end{figure}

The origin and nature of cosmic rays (CRs) with energies between $10^{15}$~eV
and the Greisen-Zatsepin-Kuzmin (GZK) cutoff at about $10^{20}$\,eV, measured by 
the HiRes~\cite{hires} and recently confirmed by the Auger~\cite{auger} experiments (Fig.~\ref{fig:sigma_CRs}, right), 
remains a central open question in high-energy astrophysics. One key to solving this question 
is the determination of the elemental composition of cosmic rays in this energy range. 
The candidate particles, ranging from protons to Fe ions, 
are measured with surface detectors that detect the ``extended air-showers'' (EAS) generated 
in the CR interactions with air nuclei when entering the Earth's atmosphere~\cite{Bluemer:2009zf}.

The determination of the primary energy and mass relies on hadronic Monte Carlo (MC) 
codes which describe the interactions of the primary cosmic-ray with air nuclei (N, O).
The bulk of the primary particle production is dominated by forward and soft 
QCD interactions, modeled commonly in Regge-Gribov+pQCD approaches 
with parameters tuned to reproduce the pre-LHC collider data ($E_{lab}\lesssim$~10$^{15}$~eV).
When extrapolated to energies around the GZK-cutoff, the current MCs predict energy and multiplicity flows 
differing by factors as large as three with significant inconsistencies in the forward region. 
The LHCf experiment~\cite{lhcf} has installed scintillator/silicon calorimeters inside
the tunnel area 140~m away of the ATLAS interaction point to detect neutral particles 
(photons, $\pi^0$, neutrons) close to the beam-rapidity ($|\eta|\gtrsim$~8.3).
Measurement of forward particle production in \pp, $p$-Pb, and Pb-Pb collisions at LHC energies 
($E_{lab} \approx 10^{17}$ eV) will provide strong constraints on these models and allow 
for more reliable determinations of the CR energy and composition at the highest energies.



\ack 
Support by the 7th EU Framework Programme contract FP7-ERG-2008-235071 is acknowledged.



\section*{References}

\begin{thebibliography}{9}

\bibitem{lhc}L.~Evans and P.~Bryant, JINST {\bf 3} (2008) S08001. 
\bibitem{atlas_tdr}G.~Aad {\it et al.}  [ATLAS Collaboration], arXiv:0901.0512.  
\bibitem{cms_tdr}A.~de Roeck (ed.) [CMS Collaboration],  J. Phys. G: Nucl. Part. Phys. {\bf 34} (2007) 995.
\bibitem{lhcb}B.~Adeva {\it et al.} [LHCb Collaboration], 
arXiv:0912.4179. 
\bibitem{alice}B.~Alessandro {\it et al.} [ALICE Collaboration], J. Phys. G: Nucl. Part. Phys. {\bf 32} (2006) 1295.
\bibitem{totem}V.~Berardi {\it et al.} [TOTEM Collaboration], CERN/LHCC-2004-002 (2004).
\bibitem{lhcf}Technical Design Report of the LHCf experiment, CERN/LHCC-2006-004 (2006).
\bibitem{moedal}Technical Design Report of the MOEDAL experiment, CERN/LHCC-2009-006 (2009).

\bibitem{Campbell:2006wx}J.~M.~Campbell, J.~W.~Huston and W.~J.~Stirling, Rept.\ Prog.\ Phys.\  {\bf 70} (2007) 89

\bibitem{cms_ichep10} G. Tonelli [CMS Collaboration], ICHEP 2010.
\bibitem{atlas_ichep10} F. Gianotti [ATLAS Collaboration], ICHEP 2010.
\bibitem{alice_ichep10} J. Schukraft [ALICE Collaboration], ICHEP 2010.
\bibitem{lhcb_ichep10} A. Golutvin [LHCb Collaboration], ICHEP 2010;
R.~Aaij {\it et al.} [LHCb Collaboration], arXiv:1009.2731. 

\bibitem{EBHGHK}Englert F and Brout R 1964 {\it Phys. Rev. Lett.} {\bf 13} 321; 
Higgs P W 1964 {\it Phys. Lett.} {\bf 12} 132; 
Guralnik G S, Hagen C R and Kibble T W B 1964 {\it Phys. Rev. Lett.} {\bf 13} 585.

\bibitem{Aglietti:2006ne} U.~Aglietti {\it et al.}, hep-ph/0612172. 
\bibitem{Djouadi:2005gi} A.~Djouadi,  Phys.\ Rept.\  {\bf 457}, 1 (2008). 

\bibitem{Barate:2003sz}R.~Barate {\it et al.} [LEP-Higgs Working Group], Phys. Lett. B {\bf 565}, 61 (2003). 
\bibitem{ichep2010} CDF and \dzero\ Collaborations, arXiv:1007.4587. 
\bibitem{Collaboration:2008ub}LEP-Tevatron-SLD Electroweak Working Group; arXiv:0811.4682. 

\bibitem{d'Enterria:2009wn}D.~d'Enterria, arXiv:0905.4307. 

\bibitem{Cheng:2007bu}H.~C.~Cheng, arXiv:0710.3407. 

\bibitem{technicolor}  S.~Weinberg, Phys.\ Rev.\  D {\bf 19} (1979) 1277;  
L.~Susskind, Phys.\ Rev.\  D {\bf 20} (1979) 2619. 


\bibitem{ADD}N.~Arkani-Hamed, S.~Dimopoulos and G.~R.~Dvali, Phys.\ Lett.\  B {\bf 429} (1998) 263. 
\bibitem{RS} L.~Randall and R.~Sundrum,  Phys.\ Rev.\ Lett.\  {\bf 83} (1999) 3370.  


\bibitem{ckmfitter} J.~Charles {\it et al.}  [CKMfitter Group], Eur.\ Phys.\ J.\  C {\bf 41} (2005) 1.

\bibitem{d'Enterria:2006su}See e.g. D.~d'Enterria,  J.\ Phys.\ G {\bf 34} (2007) S53.  


\bibitem{schaefer05}Schaefer T 2005  ``HUGS 2005 Lectures'' World Scientific  {\it Preprint} hep-ph/0509068 

\bibitem{Gelis:2010nm}F.~Gelis, E.~Iancu, J.~Jalilian-Marian and R.~Venugopalan, arXiv:1002.0333. 
\bibitem{Luzum:2008cw}M.~Luzum and P.~Romatschke, Phys.\ Rev.\  C {\bf 78} (2008) 034915 [Erratum-ibid.\  C {\bf 79} (2009) 039903]

\bibitem{adscft}J.~M.~Maldacena, Adv.\ Theor.\ Math.\ Phys.\  {\bf 2}, 231 (1998) [Int.\ J.\ Theor.\ Phys.\  {\bf 38}, 1113 (1999)];
E.~Witten,  
  Adv.\ Theor.\ Math.\ Phys.\  {\bf 2}, 505 (1998).
\bibitem{kovtun04}P.~Kovtun, D.~T.~Son and A.~O.~Starinets, Phys.\ Rev.\ Lett.\  {\bf 94}, 111601 (2005)
\bibitem{Kajantie:2006hv}K.~Kajantie, T.~Tahkokallio and J.~T.~Yee, JHEP {\bf 0701}, 019 (2007).  


\bibitem{Bourrely:2005qh}C.~Bourrely, N.~N.~Khuri, A.~Martin, J.~Soffer and T.~T.~Wu,  arXiv:hep-ph/0511135 (2005).

\bibitem{hires}R.~Abbasi {\it et al.} [HiRes Collaboration], Phys.\ Rev.\ Lett.\  {\bf 100}, 101101 (2008). 
\bibitem{auger} T.~Yamamoto  [Pierre Auger Collaboration], arXiv:0707.2638 [astro-ph].  

\bibitem{Bluemer:2009zf}J.~Bluemer, R.~Engel and J.~R.~Hoerandel, Prog.\ Part.\ Nucl.\ Phys.\  {\bf 63} (2009) 293. 
\bibitem{compete}J.~R.~Cudell {\it et al.} [COMPETE Collaboration], Phys.\ Rev.\ Lett.\  {\bf 89}, 201801 (2002).

\end{thebibliography}
\end{document}